\documentclass[fleqn,12pt,twoside]{article}
\usepackage{espcrc1}

\usepackage{graphicx}

\usepackage{longtable}

\newcommand{\AmS}{{\protect\the\textfont2
  A\kern-.1667em\lower.5ex\hbox{M}\kern-.125emS}}

\title{Reaction Rate Sensitivity of the $\gamma$--Process Path}

\author{T. Rauscher\address[MCSD]{Department of Physics and Astronomy,
University of Basel,\\
4056 Basel, Switzerland}
        \thanks{Supported by the Swiss NSF through a PROFIL professorship 
(2024-067428.01) and the grant \hbox{2000-061031.02}.}}
       
\begin{document}

\maketitle

\begin{abstract}
The location of the ($\gamma$,p)/($\gamma$,n) and
($\gamma$,$\alpha$)/($\gamma$,n) line at $\gamma$-process temperatures
is discussed, using recently published reaction rates based on global
Hauser-Feshbach calculations. The results can directly be
compared to previously published, classic $\gamma$-process
 discussions. The nuclei exhibiting the largest sensitivity to
uncertainties in nuclear structure and reaction parameters are
specified.
\end{abstract}

\section{INTRODUCTION}
Many proton-rich isotopes of naturally occurring stable nuclei cannot be
produced by neutron captures along the line of stability. The currently most
favored production mechanism for those p-isotopes is photodisintegration of
intermediate
and heavy elements at high temperatures in late evolution stages of
massive stars, the so-called $\gamma$-process \cite{wh}. Recent 
investigations have shown that
there still are considerable uncertainties in the description of nuclear
properties governing the relevant photodisintegration rates. This has 
triggered a
number of experimental efforts to directly or indirectly determine reaction 
rates
and nuclear properties for the $\gamma$-process (see, e.g.,
\cite{arngor,somorj,gyuri,pdesc} and references therein). However, many such 
investigations
focussed on nuclei in the $\gamma$-process path without considering whether the
rates involving these nuclei actually exhibit large uncertainties. In this work
the sensitivity of the location of the $\gamma$-process path on reaction rates
is investigated, showing which nuclei should be preferred in experimental
studies.

\section{CALCULATIONS}
This work makes use of the standard rate set (based on FRDM)
\cite{rath00} which is also used in many stellar models, e.g.\
\cite{rau02}. Similar to Table 2 in \cite{wh} for $T_9=2.5$, the resulting
branching points in the photodisintegration path are shown in 
the last 3 columns of Table \ref{tab:nice},
for three temperatures $T_9=2.0$, 2.5, 3.0. Following
the definition in \cite{wh}, for
each isotopic chain with charge number $Z$ the neutron number $N$ is
specified at which the condition $\lambda_{\gamma
\mathrm{p}}+\lambda_{\gamma \alpha}>\lambda_{\gamma \mathrm{n}}$ is
fulfilled for the first time when following an isotopic chain towards
decreasing $N$. The branching type is indicated by subscripts.

Usually, experimental investigations primarily focus on nuclei close to
these branch points. However, they should rather focus on rates which
are sensitive to the nuclear input, i.e. nuclei for which
$\lambda_{\gamma \mathrm{n}}$, $\lambda_{\gamma \mathrm{p}}$, and
$\lambda_{\gamma \alpha}$ are close. To this end, Table \ref{tab:nice}
also shows the nuclei for which $\lambda_{\gamma \mathrm{p}}$ and
$\lambda_{\gamma \alpha}$ are within factors $f \leq 3$ or $3<f\leq 10$,
respectively, of
the $\lambda_{\gamma \mathrm{n}}$ rate. Subscripts indicate which rate
is close to $\lambda_{\gamma \mathrm{n}}$. Two subscripts indicate that
$\lambda_{\gamma \mathrm{p}}$ or $\lambda_{\gamma \alpha}$ are within
the quoted range but that they are also within a factor of 3 of each
other. Note that according to the definition of the factors
a nucleus is not repetitively given in the factor 10
columns when it has already been included in the factor 3 column at the
same temperature.

\section{DISCUSSION}
A direct comparison with Table 2 of \cite{wh} shows remarkable agreement
with a few exceptions. This is surprising insofar as the previous
rate predictions made use of a number of simplifying assumptions, such
as equivalent square well potentials in the particle channels and
total neglection of excited states. The exceptions are Ba, W, Au, Hg
where the new branching points are shifted by 2 units to the more
neutron-rich side, Pb which is shifted by one unit, and Ce, Gd, Ho,
which have become more neutron-deficient by 2 neutrons. Only the
branching in Tl has been shifted by a larger amount, the branching point
has 4 neutrons less than previously. The branching type was impacted
even less: a combined $\gamma \mathrm{p}+\gamma \alpha$ branching was
changed into a pure $\gamma \alpha$ one in Ba and Au, and a $\gamma
\mathrm{p}$ one has become a combined $\gamma \mathrm{p}+\gamma \alpha$ 
branching in Ta. Incidentally, almost all impacted nuclei are within the
mass range $125\leq A\leq 150$ and $168\leq A\leq 200$ where
$\gamma$-process nucleosynthesis consistent with solar
p-abundances was found using the new rates \cite{rau02}, thus
underlining the improvement of the rate predictions.

As pointed out above, experiments targeting the sensitive rates given in
Table \ref{tab:nice} will have direct impact on $\gamma$-process
nucleosynthesis. Among them, sensitive rates at branching points
(coinciding with the nuclei given in the last 3 columns) will be the
most important.

In recent investigations checking theoretical rates against newly
measured ones it has become apparent that the largest problem
is in determining optical $\alpha$+nucleus potentials at low energies
(see \cite{somorj,gyuri,pdesc} and references therein).
Thus, the $\lambda_{\gamma \alpha}$ rates bear the largest inherent
uncertainty whereas $\lambda_{\gamma \mathrm{n}}$ and $\lambda_{\gamma
\mathrm{p}}$ have been found generally well predicted, with a few
exceptions \cite{gyuri}.
A more detailed study of the branching point sensitivity, also exploring 
a possible modification of $\gamma$-process nucleosynthesis and the
impact of different
$\alpha$+nucleus potentials, will be published elsewhere in an extended
paper.

\tabcolsep3pt
\begin{longtable}{lp{1.7cm}p{1.7cm}p{1.7cm}p{1.7cm}p{1.7cm}p{1.7cm}rrr}
\caption{Branch points and nuclei with large rate uncertainties (see
text). The subscripts at each neutron number indicate which rate
($\lambda_{\gamma \mathrm{p}}$ or $\lambda_{\gamma \alpha}$) is close 
to the $\lambda_{\gamma \mathrm{n}}$ rate
within the given factor, or the branching type, respectively.\label{tab:nice}}\\
\hline&\multicolumn{3}{c}{$\pm$ factor 3}&\multicolumn{3}{c}{$\pm$ factor 10}&\multicolumn{3}{c}{branch point}\\
\multicolumn{1}{c}{$Z$}&\multicolumn{1}{l}{2.0}&\multicolumn{1}{l}{2.5}&\multicolumn{1}{l}{3.0}&\multicolumn{1}{l}{2.0}&\multicolumn{1}{l}{2.5}&\multicolumn{1}{l}{3.0}&\multicolumn{1}{l}{2.0}&\multicolumn{1}{l}{2.5}&\multicolumn{1}{l}{3.0}\\
\hline
\endfirsthead
\caption{(Continued)}\\
\hline&\multicolumn{3}{c}{$\pm$ factor 3}&\multicolumn{3}{c}{$\pm$ factor 10}&\multicolumn{3}{c}{branch point}\\
\multicolumn{1}{c}{$Z$}&\multicolumn{1}{l}{2.0}&\multicolumn{1}{l}{2.5}&\multicolumn{1}{l}{3.0}&\multicolumn{1}{l}{2.0}&\multicolumn{1}{l}{2.5}&\multicolumn{1}{l}{3.0}&\multicolumn{1}{l}{2.0}&\multicolumn{1}{l}{2.5}&\multicolumn{1}{l}{3.0}\\
\hline
\endhead
\hline
\endfoot
 34 &    $_{ }$ &    $_{ }$ &    $_{ }$ &  42$_{\alpha}$ &    $_{ }$ &    $_{ }$ &  40$_{\alpha}$ &  40$_{\alpha}$ &  40$_{p,\alpha}$ \\
 35 &  46$_{p}$ &  46$_{p}$ &    $_{ }$ &    $_{ }$ &    $_{ }$ &  46$_{p}$ &  46$_{p}$ &  44$_{p}$ &  44$_{p}$ \\
 36 &  44$_{p,\alpha}$ &  44$_{p}$ &    $_{ }$ &  41$_{p}$ &  41$_{p}$ &  44$_{p}$ &  44$_{p,\alpha}$ &  42$_{p}$ &  42$_{p}$ \\
 37 &    $_{ }$ &  48$_{p}$ &  45$_{p}$, 48$_{p}$ &    $_{ }$ &  45$_{p}$ &    $_{ }$ &  48$_{p}$ &  48$_{p}$ &  46$_{p}$ \\
 38 &  43$_{p}$ &  43$_{p}$, 46$_{p}$ &  46$_{p}$ &  46$_{p}$ &    $_{ }$ &  43$_{p}$ &  46$_{p}$ &  46$_{p}$ &  44$_{p}$ \\
 39 &  49$_{p}$ &  49$_{p}$ &  49$_{p}$ &    $_{ }$ &    $_{ }$ &    $_{ }$ &  50$_{p}$ &  50$_{p}$ &  50$_{p}$ \\
 40 &  47$_{p}$ &  50$_{p}$ &  50$_{p}$ &    $_{ }$ &  47$_{p}$ &  47$_{p}$ &  50$_{p}$ &  50$_{p}$ &  48$_{p}$ \\
 41 &    $_{ }$ &  46$_{p}$ &    $_{ }$ &    $_{ }$ &    $_{ }$ &    $_{ }$ &  50$_{p}$ &  50$_{p}$ &  50$_{p}$ \\
 42 &  52$_{\alpha}$ &    $_{ }$ &    $_{ }$ &    $_{ }$ &  52$_{\alpha}$ &  49$_{p}$ &  52$_{\alpha}$ &  50$_{p}$ &  50$_{p}$ \\
 43 &    $_{ }$ &  54$_{p}$ &    $_{ }$ &  54$_{p}$ &    $_{ }$ &  51$_{p}$,
54$_{p}$ &  54$_{p}$ &  52$_{p}$ &  52$_{p}$ \\
 44 &  51$_{p}$, 54$_{\alpha}$ &  51$_{p}$ &  52$_{p,\alpha}$ &    $_{ }$ &  52$_{\alpha}$ &  51$_{p}$ &  54$_{\alpha}$ &  52$_{\alpha}$ &  52$_{p,\alpha}$ \\
 45 &    $_{ }$ &    $_{ }$ &    $_{ }$ &  55$_{p}$ &    $_{ }$ &  53$_{p}$,
56$_{p}$ &  56$_{p}$ &  56$_{p}$ &  56$_{p}$ \\
 46 &    $_{ }$ &  53$_{\alpha}$ &    $_{ }$ &  53$_{\alpha}$,
56$_{\alpha}$ &  56$_{\alpha}$ &  53$_{p,\alpha}$, 54$_{p,\alpha}$ &  56$_{\alpha}$ &  54$_{\alpha}$ &  54$_{p,\alpha}$ \\
 47 &  57$_{p}$, 60$_{p}$ &    $_{ }$ &    $_{ }$ &    $_{ }$ &  57$_{p}$,
60$_{p}$ &  58$_{p}$, 60$_{p}$ &  58$_{p}$ &  58$_{p}$ &  58$_{p}$ \\
 48 &    $_{ }$ &  55$_{p,\alpha}$, 58$_{\alpha}$ &  55$_{p}$ &    $_{ }$ &    $_{ }$ &  54$_{\alpha}$, 58$_{\alpha}$ &  58$_{\alpha}$ &  58$_{\alpha}$ &  56$_{p}$ \\
 49 &    $_{ }$ &  59$_{p}$, 62$_{p}$ &  59$_{p}$, 62$_{p}$ &  59$_{p}$ &    $_{ }$ &    $_{ }$ &  62$_{p}$ &  62$_{p}$ &  60$_{p}$ \\
 50 &  59$_{p,\alpha}$, 62$_{\alpha}$ &    $_{ }$ &    $_{ }$ &    $_{ }$ &  59$_{p}$ &  59$_{p}$, 60$_{p}$ &  62$_{\alpha}$ &  60$_{p,\alpha}$ &  60$_{p}$ \\
 51 &  62$_{\alpha}$ &  68$_{p}$ &  63$_{p}$, 68$_{p}$ &  65$_{p}$ &  63$_{p}$ &    $_{ }$ &  68$_{p}$ &  68$_{p}$ &  66$_{p}$ \\
 52 &  65$_{\alpha}$, 70$_{\alpha}$ &    $_{ }$ &  68$_{\alpha}$ &    $_{ }$ &  68$_{\alpha}$ &  63$_{\alpha}$ &  68$_{\alpha}$ &  68$_{\alpha}$ &  66$_{\alpha}$ \\
 53 &  67$_{p}$ &  67$_{p}$ &    $_{ }$ &  72$_{p}$ &    $_{ }$ &  67$_{p}$,
70$_{p}$ &  70$_{p}$ &  70$_{p}$ &  70$_{p}$ \\
 54 &  67$_{\alpha}$ &  70$_{\alpha}$ &  68$_{p,\alpha}$ &  72$_{\alpha}$ &    $_{ }$ &  65$_{\alpha}$, 70$_{p,\alpha}$ &  70$_{\alpha}$ &  68$_{\alpha}$ &  68$_{p,\alpha}$ \\
 55 &  71$_{p}$ &  74$_{p}$ &  74$_{p}$ &  76$_{p}$ &  71$_{p}$ &  69$_{p}$ &  74$_{p}$ &  74$_{p}$ &  72$_{p}$ \\
 56 &  69$_{\alpha}$ &    $_{ }$ &  72$_{p,\alpha}$ &    $_{ }$ &  72$_{\alpha}$, 74$_{\alpha}$ &  67$_{p}$ &  74$_{\alpha}$ &  72$_{\alpha}$ &  70$_{p,\alpha}$ \\
 57 &  73$_{p}$, 78$_{p}$ &  73$_{p}$, 78$_{p}$ &  78$_{p}$ &    $_{ }$ &    $_{ }$ &  73$_{p}$, 76$_{p}$ &  78$_{p}$ &  76$_{p}$ &  76$_{p}$ \\
 58 &  78$_{\alpha}$ &  74$_{\alpha}$ &  74$_{p,\alpha}$ &  76$_{\alpha}$ &  76$_{\alpha}$ &  71$_{p}$, 72$_{p,\alpha}$ &  76$_{\alpha}$ &  74$_{\alpha}$ &  72$_{p,\alpha}$ \\
 59 &  77$_{p}$ &    $_{ }$ &  75$_{p}$, 80$_{p}$ &  82$_{p}$,
84$_{\alpha}$ &  75$_{p}$, 77$_{p}$, 80$_{p}$, 82$_{p}$ &  77$_{p}$,
82$_{p}$ &  80$_{p}$ &  80$_{p}$ &  80$_{p}$ \\
 60 &  80$_{\alpha}$ &  78$_{p,\alpha}$ &  73$_{p}$, 76$_{p,\alpha}$,
78$_{p}$ &  75$_{\alpha}$, 84$_{\alpha}$ &  73$_{p,\alpha}$,
75$_{\alpha}$ &    $_{ }$ &  78$_{\alpha}$ &  78$_{p,\alpha}$ &  74$_{p}$ \\
 61 &  81$_{p}$ &    $_{ }$ &  79$_{p}$ &  84$_{\alpha}$ &  77$_{p}$,
79$_{p}$, 81$_{p}$ &  77$_{p}$, 81$_{p}$ &  84$_{\alpha}$ &  82$_{p}$ &  82$_{p}$ \\
 62 &  79$_{\alpha}$, 82$_{\alpha}$ &  77$_{\alpha}$ &    $_{ }$ &  86$_{\alpha}$ &  79$_{p}$, 82$_{p,\alpha}$, 84$_{\alpha}$ &  77$_{p,\alpha}$,
79$_{p}$, 80$_{p}$, 82$_{p}$ &  84$_{\alpha}$ &  80$_{p}$ &  80$_{p}$ \\
 63 &  88$_{\alpha}$ &    $_{ }$ &    $_{ }$ &  86$_{\alpha}$ &  84$_{\alpha}$ &  84$_{\alpha}$ &  88$_{\alpha}$ &  84$_{\alpha}$ &  82$_{p}$ \\
 64 &  85$_{\alpha}$, 88$_{\alpha}$ &  79$_{p}$, 81$_{p}$, 86$_{\alpha}$ &  79$_{p}$, 81$_{p}$, 85$_{\alpha}$ &  79$_{p}$ &    $_{ }$ &  77$_{\alpha}$ &  88$_{\alpha}$ &  84$_{\alpha}$ &  82$_{p}$ \\
 65 &  87$_{\alpha}$ &  88$_{\alpha}$ &  86$_{p,\alpha}$ &  90$_{\alpha}$ &  86$_{\alpha}$ &  83$_{p}$, 88$_{p,\alpha}$ &  88$_{\alpha}$ &  86$_{\alpha}$ &  84$_{p,\alpha}$ \\
 66 &  83$_{\alpha}$ &    $_{ }$ &  85$_{\alpha}$, 86$_{\alpha}$ &  87$_{\alpha}$, 90$_{\alpha}$ &  87$_{\alpha}$, 88$_{\alpha}$ &  88$_{\alpha}$ &  90$_{\alpha}$ &  88$_{\alpha}$ &  86$_{\alpha}$ \\
 67 &  90$_{p,\alpha}$, 92$_{p,\alpha}$ &  83$_{p,\alpha}$,
87$_{p,\alpha}$ &  85$_{\alpha}$ &    $_{ }$ &  90$_{p}$ &  83$_{p}$,
87$_{p}$ &  88$_{\alpha}$ &  88$_{p}$ &  88$_{p}$ \\
 68 &  89$_{\alpha}$ &  83$_{\alpha}$ &  83$_{p}$, 90$_{\alpha}$ &  91$_{\alpha}$, 94$_{\alpha}$ &  87$_{\alpha}$, 90$_{\alpha}$, 92$_{\alpha}$ &  87$_{\alpha}$, 88$_{\alpha}$ &  92$_{\alpha}$ &  90$_{\alpha}$ &  88$_{\alpha}$ \\
 69 &  91$_{\alpha}$, 96$_{\alpha}$ &  89$_{p}$, 94$_{p,\alpha}$ &  89$_{p}$,
94$_{p}$ &  89$_{\alpha}$ &    $_{ }$ &    $_{ }$ &  96$_{\alpha}$ &  92$_{p}$ &  92$_{p}$ \\
 70 &  91$_{\alpha}$ &  89$_{\alpha}$, 94$_{\alpha}$ &  92$_{p,\alpha}$ &  93$_{\alpha}$, 98$_{\alpha}$ &    $_{ }$ &  87$_{\alpha}$, 89$_{\alpha}$,
94$_{\alpha}$ &  96$_{\alpha}$ &  94$_{\alpha}$ &  92$_{p,\alpha}$ \\
 71 &  98$_{\alpha}$, 100$_{\alpha}$ &  93$_{p}$, 96$_{p}$ &  93$_{p}$,
96$_{p}$ &  95$_{\alpha}$ &  91$_{p}$ &  91$_{p}$ &  96$_{\alpha}$ &  96$_{p}$ &  94$_{p}$ \\
 72 & 102$_{\alpha}$ &  93$_{\alpha}$, 98$_{\alpha}$ &  91$_{\alpha}$,
96$_{\alpha}$ &  95$_{\alpha}$, 100$_{\alpha}$ &  96$_{\alpha}$ &  89$_{\alpha}$, 94$_{\alpha}$ & 100$_{\alpha}$ &  96$_{\alpha}$ &  94$_{\alpha}$ \\
 73 & 104$_{\alpha}$ &  95$_{p}$, 100$_{\alpha}$, 102$_{\alpha}$ &  95$_{p}$ &  97$_{\alpha}$, 99$_{\alpha}$ &    $_{ }$ &  98$_{p}$, 100$_{p,\alpha}$,
102$_{p,\alpha}$ & 102$_{\alpha}$ &  98$_{p,\alpha}$ &  98$_{p}$ \\
 74 &  99$_{\alpha}$ & 102$_{\alpha}$ &  93$_{p,\alpha}$, 98$_{\alpha}$ & 101$_{\alpha}$, 104$_{\alpha}$ &  95$_{\alpha}$, 97$_{\alpha}$, 100$_{\alpha}$ &  95$_{p,\alpha}$, 100$_{\alpha}$ & 104$_{\alpha}$ & 102$_{\alpha}$ &  98$_{\alpha}$ \\
 75 & 101$_{\alpha}$, 106$_{\alpha}$ &  99$_{p,\alpha}$ &  99$_{p}$ &    $_{ }$ &  97$_{p}$, 104$_{\alpha}$ &  97$_{p}$, 102$_{p,\alpha}$ & 106$_{\alpha}$ & 102$_{\alpha}$ & 102$_{p,\alpha}$ \\
 76 & 103$_{\alpha}$ & 101$_{\alpha}$, 106$_{\alpha}$ &  97$_{p,\alpha}$,
102$_{\alpha}$ & 108$_{\alpha}$ &  99$_{\alpha}$, 104$_{\alpha}$ &  99$_{\alpha}$, 104$_{\alpha}$ & 106$_{\alpha}$ & 104$_{\alpha}$ & 102$_{\alpha}$ \\
 77 & 105$_{\alpha}$ &    $_{ }$ & 106$_{p,\alpha}$ & 103$_{\alpha}$,
110$_{\alpha}$ & 106$_{\alpha}$, 108$_{\alpha}$ & 101$_{p}$ & 110$_{\alpha}$ & 106$_{\alpha}$ & 104$_{p}$ \\
 78 & 109$_{\alpha}$, 112$_{\alpha}$ & 105$_{\alpha}$, 108$_{\alpha}$ &    $_{ }$ & 107$_{\alpha}$, 110$_{\alpha}$ & 103$_{\alpha}$ & 101$_{\alpha}$,
103$_{\alpha}$, 106$_{\alpha}$ & 109$_{\alpha}$ & 106$_{\alpha}$ & 106$_{\alpha}$ \\
 79 &    $_{ }$ & 107$_{p,\alpha}$, 109$_{\alpha}$ & 107$_{p}$ & 111$_{\alpha}$, 112$_{\alpha}$ &    $_{ }$ & 105$_{p,\alpha}$ & 112$_{\alpha}$ & 110$_{\alpha}$ & 110$_{\alpha}$ \\
 80 &    $_{ }$ & 107$_{\alpha}$, 109$_{\alpha}$ & 105$_{\alpha}$,
108$_{\alpha}$ &    $_{ }$ & 109$_{\alpha}$, 110$_{\alpha}$ & 110$_{\alpha}$ & 110$_{\alpha}$ & 110$_{\alpha}$ & 108$_{\alpha}$ \\
 81 &    $_{ }$ &    $_{ }$ &    $_{ }$ & 112$_{\alpha}$ & 109$_{p,\alpha}$,
112$_{p}$ & 109$_{p}$, 110$_{p}$, 112$_{p}$ & 112$_{\alpha}$ & 110$_{p}$ & 110$_{p}$ \\
 82 & 111$_{\alpha}$ & 109$_{\alpha}$, 112$_{\alpha}$, 113$_{\alpha}$ & 107$_{p,\alpha}$, 110$_{\alpha}$ & 118$_{\alpha}$ & 105$_{p,\alpha}$,
107$_{\alpha}$ & 105$_{p}$, 109$_{\alpha}$, 113$_{\alpha}$ & 114$_{\alpha}$ & 113$_{\alpha}$ & 110$_{\alpha}$
\end{longtable}

\end{document}